\documentclass[universe,article,accept,moreauthors,pdftex]{Definitions/mdpi} 

\usepackage{latexsym}
\usepackage{amsmath}
\usepackage{amssymb}

\newcommand{\tr}[1]{\operatorname{Tr}\left[ {#1} \right]} 
\newcommand{\ket}[1]{\left|#1 \right \rangle\! \vphantom{\left( #1 \right)^A}} 
\newcommand{\bra}[1]{\left\langle #1 \right | \vphantom{\left(#1\right)^A}} 
\newcommand{\sys}{\mathcal{S}}  
\newcommand{\ev}{\text{event}}  

\newcommand{\rank}[1]{\relax\ifmmode\operatorname{rank}#1\else rank-$#1$\fi}

\firstpage{1} 
\pubvolume{00}
\issuenum{1}
\articlenumber{5}
\pubyear{2020}
\copyrightyear{2020}
\history{Received: 19 November 2020; Accepted: 9 December 2020; Published: date}

\Title{The 
	Montevideo Interpretation: How the Inclusion of a Quantum Gravitational Notion of Time Solves the Measurement Problem}

\Author{{Rodolfo Gambini} 
$^{1}$ and Jorge Pullin $^{2,}$*}

\AuthorNames{Rodolfo Gambini and Jorge Pullin}

\address{%
$^{1}$ \quad Instituto de Física, Facultad de Ciencias, Universidad de la República, Montevideo 11400, Uruguay; rgambini@fisica.edu.uy\\
$^{2}$ \quad Department of Physics and Astronomy, Louisiana State University,
  Baton Rouge, LA 70803-4001, USA}

\corres{Correspondence: pullin@lsu.edu}

\abstract{We review the Montevideo Interpretation of quantum mechanics, which is based on the use of real clocks to describe physics, using the framework that was recently introduced by H\"ohn, Smith, and Lock to treat the problem of time in generally covariant systems. These new methods, which solve several problems in the introduction of a notion of time in such systems, \mbox{do not} change the main results of the Montevideo Interpretation. The use of the new formalism makes the construction more general and valid for any system in a quantum generally covariant theory. We find that, as in the original formulation, a fundamental mechanism of decoherence emerges that allows for supplementing ordinary environmental decoherence and avoiding  its criticisms. \mbox{The recent} results on quantum complexity provide additional support to the type of global protocols that are used to prove that within ordinary---unitary---quantum mechanics, no definite event---an outcome to which a probability can be associated---occurs. In lieu of this, states that start in a coherent superposition of possible outcomes always remain as a superposition. We show that, if one takes into account fundamental inescapable uncertainties in measuring length and time intervals due to general relativity and quantum mechanics, the previously mentioned global protocols no longer allow for distinguishing whether the state is in a superposition or not.
One is left with a formulation of quantum mechanics purely defined in quantum mechanical terms without any reference to the classical world and with an intrinsic operational definition of quantum events that does not need external observers. }


\begin{document}

\section{Introduction}

The act of measurement plays a role in quantum mechanics that is unlike any it had in classical theories. Physical predictions change after a measurement. We take the point of view that our understanding of quantum mechanics is not complete until a well defined understanding of the measurement process is in place. We want a complete characterization of the process
$\rho \rightarrow \rho^{event}$,
which is, to~go from a coherent superposition to a state that can be attributed definite outcomes. Environmental decoherence~\cite{decoherence} is a proposal that comes close, identifying instances in which the transition occurs for all practical purposes. There are additional proposals, like quantum Darwinism~\cite{quda} and coarse-grained measurements~\cite{coarse}.

However, solutions that are based on decoherence are incomplete, since they give rise to a subjective and ill defined notion of event.
There are protocols that show  that the ‘decoherence solution’ is apparent. 
We are not the only ones levying these criticisms, and~lots of work has been devoted to different modifications and/or re-interpretations of the theory to clarify the situation.  Several involve compromises, for~example, giving up the possibility of a single-world description as a way of keeping the formalism of quantum theory intact~\cite{givingup,givingup-1,givingup-2,givingup-3}, or~losing  the notion of objectivity~\cite{objectivity,objectivity-1}, accepting makeshift modifications to the  theory in order to restore a single-world picture~\cite{singleworld,singleworld-1,singleworld-2,singleworld-3,singleworld-4,singleworld-5,singleworld-6}, or~even combinations of all the above~\cite{allof}. Here, we would like to present a realist description, with~minimal premises and a well defined notion of~event.

The Montevideo Interpretation of quantum mechanics~\cite{montevideo,montevideo-1,montevideo-2} is, in~reality, an~extension of the quantum mechanics obtained by taking into account that time must be a physical observable, and~not a classical, parameter\footnote{The idea of using a physical clock is quite old, going back to Bergmann and Dirac~\cite{DiracBergmann,DiracBergmann-1}.}. It takes seriously the notion that everything must be quantum in a quantum universe. \mbox{It does} not  invoke a classical world in order in order to define the quantum theory, like the Copenhagen interpretation does. Alternatively, it could be interpreted as confirming the preferred basis to be used in a Many Worlds interpretation~\cite{butterfield}. When considering that time is a physical observable is even more natural when one considers gravity, as~it is well known that, in~generally covariant theories, the~time coordinate is just a gauge parameter and clocks must be constructed from physical~quantities.

In this paper, we succinctly present the main results that lead to the Montevideo Interpretation incorporating recent advances in the understanding of problems that are related to it, pointing, in~each case, to~references where the analyses are more extensive, and~showing its general applicability to quantum covariant~systems.

In Section \ref{sec2}, we introduce a notion of time for generally covariant systems. Section \ref{sec3} discusses fundamental limits to space-time measurements based on general relativity and quantum mechanics. Section \ref{sec4} discusses how quantum measurements are possible when real clocks are considered. \mbox{Section \ref{sec5}} discusses the objective notion of the event that we consider. We end with~conclusions.

\section{A Quantum Notion of Time in Generally Covariant~Systems}\label{sec2}

There has been recent progress in understanding, with~precision, how to recover the usual quantum descriptions, which is, the~Schr\"odinger or Heisenberg pictures in totally covariant systems, like general relativity~\cite{hoehn,hoehn-1}. It provides a systematic analysis that cannot be ignored in any attempt to advance the understanding of the problem of time. In~their main paper regarding the problem of time, H\"ohn, Smith, and~Lock~\cite{hoehn,hoehn-1} have shown how it is possible to consistently treat three historical approaches to this problem. Here, we will review the first two approaches at a classical level in order to focus on what follows in the nature of the time parameter~used. 

One wishes to describe a totally constrained system with a constraint $C_H$, whose classical dynamics is defined by the gauge transformations that are generated by such a constraint. A~first approach can be made in terms of relational physical observables\footnote{Also known as evolving constants of the motion.}
$F_{f,T}(\tau)$ \cite{rovelli,rovelli-1,rovelli-2,dittrich,dittrich-1,dittrich-2}. They are Dirac observables, which is they commute with the constraint(s) and coincide with the function of the kinematical variables $f$ when the variable $T$ is equal to the parameter $\tau$, also belonging to the kinematical space.
A second approach can be to fix a gauge (a choice of time in the classical theory) and deparameterizing. For~that purpose, a~canonical transformation is introduced in the classical system, which gives rise to a time variable $T$ and its canonically conjugate momentum $P_T$, and~relational Dirac observables that are associated to the remaining original canonical variables
 $F_{q_i,T}(\tau)$, $F_{p_i,T}(\tau)$. The~gauge is fixed, for~instance, $T=0$ and one solves for $P_T$, which eliminates the gauge dependence. That this can always be done with ``good clocks'', as~shown in~\cite{hoehn,hoehn-1}\footnote{In this paper we assume that the procedure for identification of good clocks can be carried out for realistic quantum gravitational systems. It is clear that, at~the moment, this can only be done explicitly for limited examples.}. The~$F_{q_i,0}(\tau)$ now obey canonical equations that describe their evolution in terms of a true Hamiltonian $H_{\rm true}$ that result from the deparameterization and satisfy initial conditions $F_{q_i,0}(0)=q_i^0$ and $F_{p_i,0}(0)=p_i^0$ at $T=0$. One then recovers the classical evolution in terms of the classical clock variable $\tau$. The~third approach, which is that of Page and Wootters~\cite{pawo},  is purely quantum and cannot be treated~classically. 

In order to quantize these approaches, one needs to introduce, in~each case, a~time operator. \mbox{Time operators} are not defined in the physical space of states. The~latter are annihilated by constraints and, therefore, by~the total Hamiltonian (which is a linear combination of the constraints in a totally constrained theory). Therefore, physical operators have to commute with the total Hamiltonian and, therefore, are constants of the motion. Even at the kinematical level (where states that are not annihilated by the total Hamiltonian and the constraints), it is problematic to quantize a time variable. In~fact, as~Pauli has observed, it is not possible to define a self-adjoint operator conjugate to a bounded-below Hamiltonian. Later on, this result was extended by Unruh and Wald~\cite{unwa}, \mbox{who showed} that, if~the Hamiltonian is bounded{-below}
, there is no self-adjoint operator that only can run forward in time. In~other words any realistic clock that runs forward in time has a non vanishing probability of running backwards. Although~there are no self-adjoint operators associated to a ``time'' variable that satisfy the conditions of a good clock, one can use generalized  measurements in terms of positive operator-valued measures (POVM) and effect operators in order to describe clock readings. This was the strategy followed by H\"ohn, Smith and Lock~\cite{hoehn,hoehn-1}. However, these operators are necessarily defined in the kinematical space and do not correspond to any Dirac observable in the physical space. \mbox{One may} wonder if operators defined in the kinematical space are observable. After~all any physical system in a background independent theory should obey the physical laws derived from the constraint. Let us assume that that the clock system is an element of our Universe and, therefore, satisfies the quantum and general relativistic laws, for~instance, it is an atomic clock. The~observable time for this kind of clocks cannot be described by kinematical operators that would completely ignore the Einstein equations. Of~course, if~we come back to the physical Hilbert space, we again have to face the problem of time, given the fact that Dirac observables are constants of the~motion. 

One could ask oneself if there is anything gained with this construction. The~answer is clearly positive. The~most important result is that, following any of the mentioned techniques, i.e., introducing relational physical observables, deparameterization or the Page--Wootters formalism, one ends up with a standard formulation of quantum mechanics in totally covariant systems in the Schr\"odinger or Heisenberg picture. Like in non-relativistic quantum mechanics, time is treated differently than any other variable. In~ordinary quantum mechanics, it is a classical variable and, in~a totally constrained system, a~kinematical variable. In~a totally constrained and quantum mechanical world, such variables are idealizations that lead to the simplest description of the evolution in terms of unitary evolution operators. However, the~measurement of time is always done while using a real clock that is subject to gravitational and quantum laws. These clocks are subject to the uncertainties and fluctuations of all quantum systems and, therefore, are never exact. Our observation, that we here re-derive in the language of H\"ohn, Smith, and~Lock, is that the evolution that is described in terms of a real clock satisfies a modified master~equation.

We shall consider a physical situation described by a set of variables $q,p$ belonging to a constrained system that includes the physical system that we wish to study and a ``clock'' that will be used to keep track of the passage of time, which we will assume takes continuum values. A~simple example of a clock variable could be the position of certain free particle. The~evolving constant of the motion associated with it is $F_{T,t_c}(\tau)$\footnote{We are using the notation of H\"ohn, Smith, and~Lock. Readers that have followed our previous papers might notice that there we referred to it as the operator on the physical space $T(\tau)$.}. As~before, $T$ and $t_c$ are kinematical quantities and $\tau$ a number.  \mbox{It corresponds} to a relational physical observable that takes the value of $T$ when the parameter $\tau$ takes the value of the auxiliary kinematical time $t_c$ that will, in~the quantum theory, be modeled by a POVM. We also identify the physical variables that we wish to study as the relational physical observable $F_{O,t_c}(\tau)$ that is associated to the kinematical variable $O$. We then proceed to quantize the system by promoting the observable and clock to self-adjoint quantum operators $\hat{F}$ acting on the kinematical Hilbert space and $t_c$ is taken as an auxiliary  variable belonging to the kinematical space (or represented by a POVM in the quantum theory). We are quantizing the relational physical observables following the procedure that is discussed in~\cite{hoehn,hoehn-1}.

As we did in reference~\cite{pedagogical}, we call the clock  eigenvalues (the eigenvalues of the relational observable $F_{T,t_c}(\tau)$) $T$ and $O$ the eigenvalues of the relational observable $F_{O,t_c}(\tau)$, and~we assume that both have a continuum spectrum. They satisfy,
\begin{equation}
    {\hat F}_{T,t_c}(\tau=t_c)\vert T,k,\tau\rangle=T\vert T,k,\tau\rangle,
\end{equation}
which are eigenvalue equations in the kinematical space. The~states of such space $\vert T,k,\tau\rangle$ are labeled by the eigenvalues of ${\hat F}_{T,t_c}$, of~other parameterized Dirac observables that we collectively denote as $k$. The~observables that are associated with $T$ and $k$ are dependent on $\tau$, which is why we include it among the labels of the state. We observe that this operator may be written in terms of Dirac observables and $\tau$, and~we ``normalize'' the eigenstates in the physical state space
\begin{equation}
    \langle T,k,\tau\vert T',k,\tau\rangle_{\rm phys}=\delta(T-T'),
\end{equation}
where $\tau$ becomes a non-observable parameter.\footnote{We are assuming a discrete spectrum for the other observables, if~one had a continuous one, an~extra Dirac delta would be needed in the expression.}

 The projector in the physical space that corresponds to finding the time variable $T$ in the interval 
$[T_0-\Delta T,T_0+\Delta T]$ for a given  $\tau$ is,
\begin{equation}
P_{T_0}(\tau) =\int_{T_0-\Delta T}^{T_0+\Delta T} dT \sum_k \vert T,k,\tau>\langle T,k,\tau\vert.
\end{equation}
where $k$ denotes the eigenvalues of other Dirac observables that form a complete set with $\hat{F}_{T,t_c}(\tau)$, which we have assumed  has a continuum spectrum (this is not really necessary, which is why we kept a sum over it rather than an integral). Similarly, the~projector for O is
\begin{equation}
P_{O_0}(\tau) =\int_{O_0-\Delta O}^{O_0+\Delta O} dO \sum_j \vert O,j,\tau\rangle\langle O,j,\tau\vert.
\end{equation}

In order to determine the simultaneous probability of finding $O$ in the interval $\Delta O$ around $O_0$ and the clock around time $T_0$, we start by assuming that the measurements occurs when the parameter takes the value $\tau$,
\begin{equation}
{\cal P}(O\in[O_0-\Delta O,O_0+\Delta O], T\in [T_0-\Delta T,T_0+\Delta T],\tau)={{\rm Tr}(P_{O_0}(\tau)P_{T_0}(\tau)\rho P_{T_0}(\tau))},
\end{equation}
where $\rho$ is a density matrix in the physical space of states and we used the cyclic property of the trace and that $P_{O_0}$ is a projector and, therefore, equal to its square.
We can compute, in~an analogous way, the~probability of finding $T$ in the given interval. The~conditional probability of finding the observable in its interval when the clock is in its own interval would be given by the ratio of the simultaneous probability and the probability of finding $T$ in the given interval. 
Taking into account that we have a complete ignorance of the value of $\tau$, and~that the above equations depend on $\tau$, we need to take the average of these expressions in all possible values of $\tau$. To~do that, we consider all of the values of $\tau$ leading to a non-vanishing value of ${\rm Tr}(P_{T_0}(\tau)\rho)$, let us call ${\tau_1}^{T_0}$ the minimum of these values and ${\tau_2}^{T_0}$ the maximum, which is $\tau \in [{\tau_1}^{T_0},{\tau_2}^{T_0}]$ and $\Delta t^{T_0}=\tau_2^{T_0}-{\tau_1}^{T_0}$. Subsequently, by~taking the average on the possible values of the unobservable parameter and simplifying the factors $1/\Delta t^{T_0}$ in numerator and denominator, we get, for~the conditional probability,
\begin{equation}\label{conditional}
  {\cal P}(O\in[O_0-\Delta O,O_0+\Delta O]\vert T\in [T_0-\Delta T,T_0+\Delta T])=\frac{\int_{{\tau_1}^{T_0}}^{{\tau_2}^{T_0}}
 d\tau {\rm Tr}(P_{O_0}(\tau)P_{T_0}(\tau)\rho P_{T_0}(\tau))}{\int_{{\tau_1}^{T_0}}^{{\tau_2}^{T_0}}d\tau {\rm Tr}(P_{T_0}(\tau)\rho)}.
 \end{equation}
{We integrate} 
 the numerator and denominator separately, since the numerator with its integral corresponds, up~to a factor $1/\Delta t^{T_0}$, to~the joint probability of $O_0,T_0$, and~the integral of the, up~to the same factor, to~the probability of $T_0$.
The reason for this averaging is that we do not know for which value of the kinematical time $\tau$ the clock took the value $T_0$. Notice that something similar occurs if one deparameterizes, because, in~this case, the~parameter $\tau$ would correspond to a classical variable. If~we could have a perfect clock, such that, to~each $\tau$, corresponded only one $T$, the~expression would reduce to the one of ordinary quantum mechanics. This behavior is still true if one considers simple theories, like in ordinary quantum mechanics, it does not depend on the fact that we are working with a totally constrained theory. It is enough to take into account that the Schr\"odinger time is an idealized variable about which one does not have perfect~information.

Let us assume that we have chosen a physical clock whose interactions with the system of interest are negligible and that behaves semiclassically with small quantum fluctuations. It will typically be an atomic clock that is based on a periodic system with small deviations from the ideal non observable time $\tau$. Therefore, in~first approximation, one expects to recover the ordinary Schr\"odinger evolution plus small corrections. Let us then assume that we can divide the density matrix of the whole system into a product form between clock and system, $\rho=\rho_{cl}\otimes \rho_{sys}$, and~the evolution will be given by a unitary operator that is also of product type $U = U_{cl} \otimes U_{sys}$. 

As shown by H\"ohn, Smith, and~Lock~\cite{hoehn,hoehn-1}, the~evolution in terms of a density matrix at time $\tau$ is given by the usual probability of measuring the value of $O$ at a time $\tau$, in~a state $\rho$,
\begin{equation}
  {\cal P}\left(O\vert\tau\right)\equiv \frac{{\rm Tr}\left(P_{O}(0)\rho(\tau)\right)} {{\rm Tr}\left(\rho(\tau)\right)}\label{ordinary}
\end{equation}

Because $\tau$ is unknown, we would like to shift to a description where we have density matrices as functions of the observable time T. To~do this, we start from the conditional probability (\ref{conditional}), and~make the separation between clock and system explicit,
\begingroup\makeatletter\def\f@size{8}\check@mathfonts
\def\maketag@@@#1{\hbox{\m@th\normalsize\normalfont#1}}%
\begin{eqnarray} \label{condprob2}
{\cal P}\left( O\in [O_0\pm\Delta O]|T \in 
[T_0\pm\Delta T]\right) &=&
\frac{\int_{t_1^{T_0}}^{t_2^{T_0}} d\tau\,{\rm Tr}\left(U_{\rm sys}(\tau)^\dagger P_O(0)
U_{\rm sys}(\tau) U_{\rm cl}(\tau)^\dagger P_T(0) U_{\rm cl}(\tau)\rho_{\rm sys}\otimes 
\rho_{\rm cl}\right) }
{{\int_{t_1^{T_0}}^{t_2^{T_0}} d\tau\,{\rm Tr}\left(P_T(\tau)\rho_{\rm cl} \right) {\rm Tr}\left(\rho_{\rm sys}\right)}}\nonumber\\
&=&\frac
{{\int_{t_1^{T_0}}^{t_2^{T_0}} d\tau\,{\rm Tr}\left(U_{\rm sys}(\tau)^\dagger P_O(0)
U_{\rm sys}(\tau) \rho_{\rm sys}\right){\rm Tr}\left(U_{\rm cl}(\tau)^\dagger P_T(0) U_{\rm cl}(\tau) 
\rho_{\rm cl}\right)}}
{{\int_{-t_1^{T_0}}^{t_2^{T_0}} d\tau\,
{\rm Tr}\left(P_T(\tau)\rho_{\rm cl} \right) {\rm Tr}\left(\rho_{\rm sys}\right)}}.
\end{eqnarray}
\endgroup

{By introducing} 
the probability that the measurement of T has occurred at the (unknown) time $\tau$
\begin{equation}
{\cal P}_\tau(T) \equiv 
\frac{{\rm Tr}\left(P_T(0) U_{\rm cl}(\tau)\rho_{\rm cl} U_{\rm cl}(\tau)^\dagger\right)}
{\int_{t_1^{T}}^{t_2^{T}}d\tau\,{\rm Tr}\left(P_T(\tau) \rho_{\rm cl}\right)},
\end{equation}

Additionally, noticing from the above expression that $\int_{{\tau_1}^T}^{{\tau_2}^T}d\tau{\cal P}_\tau(T)=1$, we may introduce a $T$ dependent density matrix of the system at time $T$
\begin{equation}
\rho(T)\equiv\int_{{\tau_1}^T}^{{\tau_2}^T}d\tau{\cal P}_\tau(T)U_{sys}(\tau){\rho_{sys}}U_{sys}^\dagger(\tau).
\end{equation}

Additionally, noting that ${\rm Tr}(\rho(T))={\rm Tr}(\rho_{sys})$ one can derive from (5) the ordinary expression for                         the probabilities in quantum mechanics given in (4) but with an effective density matrix given by $\rho(T)$. Because~one ends up with a superposition of density matrices evolved unitarily for different values of $\tau$, the~effective evolution of the physical density matrix $\rho(T)$ is not~unitary.

In reference~\cite{pedagogical}, we have shown that, if~the real clock behaves semi-classically and we assume that ${\cal P}_t(T) = f(T-t)$ is a peaked symmetric function approximated by a Dirac delta, which is $f(T-t) =\delta(T-t)+b(T)\delta''(T-t)+\ldots$ with a width $b(T)$ that grows with time, one gets
\begin{equation}
\frac{\partial \rho(T)}{ \partial T} =i [\rho(T),H] +\sigma(T) [H,[H,\rho(T)]]+\ldots
\end{equation}
where $\sigma(T)=\partial b(T)/{\partial T}$ is the rate of spread of the width of the clock state. This is an equation of the Lindblad~\cite{lindblad} type that should be considered to be the master equation that describes the actual evolution of any physical system evolving in terms of a real clock. The~correcting term is proportional to $\sigma(T)$, a~quantity that depends on the particular clock used  controls the decoherence induced by the use of physical clocks on the evolution of the states.  As~we shall see, there are fundamental bounds of how well a clock can be and for the value of $b(T)$ that are implied by this bound; we will recover a master equation for optimal physical~clocks.

Summarizing: \emph{{Schr\"odinger's equation is only approximate, it is a description in terms of a classical time that is not accessible to observers in a quantum and covariant universe. When one takes into account that clocks are physical systems just like any other and that the universe has covariant and quantum laws, the~evolution needs to be modified. It ends up being described by a master equation that includes additional effects of loss of coherence. The~origin of the lack of unitarity is the fact that definite statistical predictions are only possible by repeating an experiment. If~one uses a real clock, which has thermal and quantum fluctuations, each experimental run will correspond to a different value of the evolution parameter. Therefore, the~statistical prediction will correspond to an average over several intervals and, therefore, its evolution cannot be unitary.}}

\section{Fundamental Limits to Space-Time~Measurements}\label{sec3}

From the above analysis, it is clear that, if~there are fundamental limitations of how good a clock can be, then the use of real clocks will introduce an additional, fundamental, source of decoherence. 
Because we do not have a complete theory of quantum gravity, this is still a contentious issue. Phenomenological arguments have been given by Salecker and Wigner, Karolyhazy, Ng and van Dam, Amelino-Camelia, Ng and Lloyd, and~Frenkel~\cite{several,several-1,several-2,several-3,several-4,several-5,several-6,several-7}, leading  to similar estimations that are based on two main effects: quantum fluctuations and black hole formation. We have recently given a simple argument leading to a fundamental minimum uncertainty in the determination of time intervals consistent with the previous estimations. It only relies in the uncertainty principle and time dilation in a gravitational field~\cite{bounds}.
Schematically, the~argument is as follows: let us consider a microscopic quantum system playing the role of a clock and a macroscopic observer that interacts with the clock interchanging signals. Let us start by considering the time-energy uncertainty relation,
$\Delta E \Delta t_c \ge \hbar$, 
where $\Delta t_c$ is of the order of the period of oscillation of the system being considered, and~$E$ is the energy of the quantum oscillator. One can consider that the macroscopic system is at an infinite (macroscopic) distance from the microscopic quantum oscillator. 
We now consider the relationship between the time that is measured by the clock locally, $t_c$, and~an observer at an infinite distance from it, $t$. The~gravitational time dilation measures the difference in the passage of proper time at different positions, as~described by a the metric tensor of space-time. It is given by
$$t=\frac{t_c}{\sqrt{1-r_S/r}}$$
where $r_S$ is the Schwarzschild radius that is given by  $r_S=2GE/c^4$

We will concentrate in the uncertainty of the observed period of oscillations. While using the standard technique for the propagation of errors of a measurement,
\begin{equation}\label{4}
  \left(\Delta t\right)^2 = \frac{1}{4}\frac{t_c^2 \left(\Delta
      r_S\right)^2}{\left(1-\frac{r_S}{r}\right)^3 r^2}+\frac{\left(\Delta t_c\right)^2}{{1-\frac{r_S}{r}}},
\end{equation}
and taking into account the definition of the Schwarzschild radius
and the time energy uncertainty relation, we get
\begin{equation}
\left(\Delta t\right)^2\ge\frac{t_c^2 G^2
  \hbar^2}{\left(1-\frac{r_S}{r}\right)^3 r^2 c^8 \Delta
  t_c^2}+\frac{\left(\Delta t_c\right)^2}{1-\frac{r_S}{r}}.
\end{equation}
 
 Recalling that $1-r_S/r$  is a positive quantity that is less than one, since the size of the clock cannot be smaller than its Schwarzschild radius, and~translating $t_c$ to $t$, one has that,
\begin{equation}\label{7}
\left(\Delta t\right)^2 \ge \frac{t_c^2 G^2 \hbar^2}{\left(1-\frac{r_S}{r}\right)^3 r^2 c^8
  \left(\Delta t_c\right)^2} + \frac{\left(\Delta
    t_c\right)^2}{1-\frac{r_S}{r}}>\frac{t^2 G^2 \hbar^2}{r^2 c^8
    \left(\Delta t_c\right)^2}+\left(\Delta t_c\right)^2.
\end{equation}

Assuming that the clock has size r and that the oscillation within it takes place at the mean speed $v$, we have that $2\pi r = v \Delta t_c$ (this actually holds in curved space~\cite{curved}) and computing the minimum of $\Delta t$ as a function of $\Delta t_c$ while using Equation~(\ref{7}) and taking into account that $v<c$,  we get,
\begin{equation}
  \label{eq:4}
  \Delta t > \sqrt{3} \pi^{1/3} t^{1/3} t_{\rm Planck}^{2/3}.
\end{equation}

{This is} 
a 
bound that agrees with those that are derived by other means by the authors mentioned above~\cite{several,several-1,several-2,several-3,several-4,several-5,several-6,several-7}. Similar fundamental uncertainties hold for spatial intervals $l$ and for any relativistic invariant interval $s$ \cite{spatial}.

If the best accuracy one can get with a clock is the one given above, it will induce via the master equation the decay of the out of diagonal terms of the density matrix $\rho(T)$ ($T$ would correspond to the $t$ of this section).
\begin{equation}
\label{16}
\rho(T)_{mn}=\rho_{mn}(0)\exp(-i\omega_{mn}T)\exp(-\omega_{mn}^2t_{Planck}^{4/3}T^{2/3})
\end{equation}

Therefore, pure states evolve approaching statistical mixtures, also known as classical mixtures, which suffer an irreversible evolution, and~the system presents a fundamental loss of coherence due to this effect. This is a fundamental effect; any physical system will lose unitarity through its~evolution.

Summarizing: \emph{{the precision with which time lapses or spatial distances can be measured is limited by quantum and gravitational effects. As~a consequence of this limitation the quantum states exhibit small deviations from the Schr\"odinger evolution. The~evolution described by real clocks becomes irreversible and exhibits a new form of loss of coherence independent from environmental decoherence implying that pure states approach classical mixtures as time evolves.}}

\section{Quantum Measurements with Real~Clocks}\label{sec4}

According with the orthodox view of the quantum measurement problem,
 the collapse of the wave-packet during measurements refers to an irreducibly indeterministic change in the state of a quantum system, contravening the deterministic and continuous evolution that was prescribed by the Schr\"odinger equation. One needs to address several questions. In~particular: under exactly what conditions does the collapse occur? In other words, the~problem of measurement in quantum mechanics arises in standard treatments as the requirement of a reduction process when a measurement takes place. Such a process is not contained within the unitary evolution of the quantum theory, but~it has to be postulated externally and is not unitary. Processes that occur during measurements are usually  justified through the interaction with a large, classical measuring device and an environment with many degrees of freedom, which is through ambient decoherence.\footnote{This obviously depends on the interpretation considered. If~one considers informational interpretations the collapse is just an update of the wavefunction and no mention of the environment is involved. We are concentrating on realist interpretations.}
 
 Objections have been levied onto two aspects of the solution of the problem of measurement through decoherence.
(1) Because the evolution of the system plus environment is unitary, the~coherence still persists and it could potentially be regained.
(2) In a picture where evolution is  unitary ``nothing ever occurs''. This is Bell’s “and/or” problem. The~final reduced  density matrix of the system plus the measurement device will describe a set of coexistent options and not alternative options with definite probabilities. To~put this feature vividly, in~terms of Schrödinger's cat: at the end of the decoherence process, the~quantum state still describes two coexistent cats, one alive and one dead. We shall argue that these two problems are related with the issue of time and then propose a solution for~them.

A first approach that we took to the analysis of the first issue is model dependent
~\cite{GaGaPu}. It consists in considering a model where the quantum system, the~measurement apparatus, and~the environment are completely under control and study its evolution in terms of real quantum clocks. One can show that with the modified evolution described by the master equation the first objection to decoherence does not apply. We will concentrate in this section in the first objection: \emph{{the quantum coherence is still there}}.
Although a quantum system interacting with an environment with many  degrees of freedom will very likely give the appearance that the initial quantum coherence of the system is lost---the density matrix of the measurement device is almost diagonal---the information regarding the original superposition could be recovered, for~instance, carrying out a measurement  that includes the environment. The~fact that such measurements are hard to carry out in practice does not prevent the issue from existing as a conceptual problem. The~persistence of correlations also manifests in closed systems in the problem of revivals:  after a very long time, the~off-diagonal terms in the reduced matrix of the system plus the measurement device become large again. So that, whatever definiteness of the observed preferred quantity that had been gained by the end of the measurement interaction, it turns out, in~the very long run, to~have been but a temporary victory. The~superposition of different outcomes reappear in the state of the measuring apparatus. This is called the problem of revivals 
(or ‘recurrence of coherence’, or~‘recoherence’). The~fundamental irreversible decoherence that is induced by the use of real clocks allows for showing that the modified evolution prevents revivals. When the multiperiodic functions in the coherences of the process induced by the environment tend to take again the original value after a Poincaré time of recurrence, the~exponential decay of the out of diagonal terms of the density matrix in Equation~(\ref{16}) for sufficiently large systems completely hides the revival under the noise amplitude~\cite{GaGaPu}.

Let us discuss the possibility of recovering the information that, after~the enviromental decoherence, lays in the environment interference terms. That is, let us try to establish whether the system remains in a coherent superposition or has become a statistical mixture. Because~the information in question is a characteristic of the total system (including the environment) being in a coherent superposition, it can be, in~principle, revealed by measuring a suitable quantity of the total system. A~typical procedure could be measuring an observable of the total system that takes different values for a coherent superposition and a statistical mixture. For~instance, that vanishes in the last case. This was proposed, for~instance, by~d'Espagnat~\cite{despagnat}. In~reference~\cite{GaGaPu}, we showed that, due to the fundamental decoherence induced by quantum clocks, the~expectation value of such observables exponentially decreases and it is increasingly difficult to distinguish from the vanishing value that results from an exact statistical mixture. The~analysis was based in a particular model of a spin interacting with a spin bath~environment.

In what follows, we are going to show that, due to the combined effect of fundamental decoherence and the bounds in the precision of the measurements of length and time intervals, this analysis may be extended to a wide range of systems~\cite{GaGaPu18} and show that it is impossible to distinguish between the expectation value for the evolved initial state and a statistical mixture not only for all practical purposes (FAPP), but~fundamentally.
The solution may be applied to a general class of global protocols that apply to any decoherence model. In~this way, we provide a criterion that works in much more general settings than a particular model. This analysis also permits providing estimates for bounds on the time in which the event~occurs.

Let us first consider a system with unitary evolution that interacts with its environment and presents ambient decoherence, then \emph{{it is in principle always possible to distinguish whether the system is in a superposition or a statistical mixture}}. An~obvious way of showing that is the following.
Let us assume that we are interested in the measurement of the observable $A$ with eigenvalues $a_i$ and eigenstates $\vert\varphi_i\rangle$, and~we are interested in comparing the evolution during the measurement process of a pure state, $$\rho_1(0)=\vert\psi\rangle\langle\psi\vert=\sum_{i,j}\alpha_i\alpha^*_j\vert\varphi_i\rangle\langle\varphi_j\vert,$$
and a statistical mixture ,
 $$\rho_2(0)=\sum_{i}\vert\alpha_i\vert^2\vert\varphi_i\rangle\langle\varphi_i\vert.$$
 
Measurements on both states lead to the same set of results with the same probabilities. However, while after a measurement outcome is observed $\rho_1$ becomes $\rho_2$, the~latter remains invariant. After~an event\footnote{A measurement with an outcome that is not necessarily registered, and~does not actualize the information in the state.} occurs, the~system becomes $\rho_2$. 
However, if~one studies the unitary evolution of the systems that are coupled to the environment, even after the interaction with the environment of the measuring device, the~total system resulting from the evolution of the first state differs from the total system resulting from the second. Without~projections breaking the unitarity, events would not occur. The~distinction between them may become harder after evolution, but~it is always possible provided that unitarity is preserved. A~measurement on the first system with a definite outcome would always require breaking the unitarity. It is usually argued by decoherentists that, at~a certain moment, the~evolution becomes ``effectively~irreversible''. 

Aaronson, Atia, and~Susskind~\cite{AaAtSu} have shown that observing the interference terms in a state as $\rho_1(t)$ is as difficult as reproducing the initial state. In~terms of Schr\"odinger's cat, measuring the interference terms is as difficult as bringing the dead cat back to life. A~trivial example of a protocol allowing for distinguishing $\rho_1(t)$ and $\rho_2(t)$ is time reversal. In~fact, we would recover the initial states, and~they are easily distinguishable. Implementing this protocol with enough approximation to distinguish the two situations in an experiment is, without~a doubt, an~extremely hard task, which~would require control over the huge number of degrees of freedom in the environment (see, however, {ref.}
~\cite{navascues} for some progress in that direction). However, the~fact that this possibility, in~principle, exists~is already an insurmountable obstacle to constructing an objective notion of event within unitary quantum mechanics at a conceptual~level.

In reference~\cite{GaGaPu18}, we have analyzed the effect of the fundamental decoherence that is induced by the use of real clocks.  If~the same time reversal is considered in this case, the~irreversibility of the evolution that is described by the master equation does not allow for recovering the initial state. In~fact, if~we consider the evolution of $\rho_1(0)$ from the initial time $0$ to $t$ and then back-reversed to $t=0$, then~we do not re obtain  $\rho_1(0)$, but~a state that differs very little from $\rho_2(0)$.

In order to quantify this difference, let us consider the observable $O=O_S\otimes I_E$, where $S$ refers to the system
and $E$ the environment. Subsequently, if~$O_S=\vert\varphi_j\rangle\langle\varphi_k\vert+\vert\varphi_k\rangle\langle\varphi_j\vert$  
the expectation values for $\rho_1(0)$ and $\rho_2(0)$ initially are
\begin{equation}
\label{eq:expectationOunitary}
\mathrm{Tr}\left(O \rho_{S,E,1}(0)\right)= \alpha_j\alpha_k^* + \alpha_k\alpha_j^*,
\end{equation}
while
\begin{equation}
\mathrm{Tr}\left(O \rho_{S,E,2}(0)\right)  = 0,
\end{equation}
where the trace is taken over the whole system, including the environment.
After evolving a time $T$ and time-reversing the system until
$T_f=2T$,
\begin{align}
\label{eq:observabledecay}
&\tr{O \big( \rho_{S,E,1}(T_f)  -\rho_{S,E,2}(T_f)  \big) } \nonumber \\
&  \lesssim \frac{\sqrt{2}\tau_D}{\sqrt{\pi} 2^{1/3} T_{\rm Planck}^{2/3}T^{1/3}}     \tr{O_\sys \big( \rho_{S,1 }(0)  -\rho_{S,2}(0)  \big) },
\end{align}
where $\tau_D$ is the usual time of environmental decoherence of the system that is coupled to the measuring device and, in~the right hand side, the~trace is taken over $S$. This shows that using the global protocol to distinguish the state evolved with the master equation from the state in the case an event occurs, becomes increasingly harder when real clocks are used and time uncertainties are taken into account. The~state evolved with the master equation becomes increasingly similar to the case in which an event occurs and it can be interpreted as a classical~mixture. 

One could consider that, after~all, even though, for~all practical purposes, both of the states are practically identical, they are still fundamentally different.
We now show that this is not the case when one takes into account that the uncertainties in time intervals and length measurements put limitations in how the states of a system is prepared or measured. We illustrate it in the paradigmatic case of a particle in a coherent superposition over two spatial locations (for more details, see~\cite{GaGaPu18}). To~do that, we take $\vert\psi\rangle=a\vert\varphi_1\rangle+b\vert\varphi_2\rangle$, where $\vert\varphi_2\rangle$ and $\vert\varphi_1\rangle$ are states that are localized at different points, and~$$\rho_1(0)=\vert\psi\rangle\langle\psi\vert,$$
with
\begin{align}
\langle x \ket{\varphi_1} &= \frac{1}{(2\pi\sigma^2)^{1/4}} \exp\left(-\frac{(x-L/2)^2}{4\sigma^2}\right), \\
\langle x \ket{\varphi_2} &= \frac{1}{(2\pi\sigma^2)^{1/4}} \exp\left(-\frac{(x+L/2)^2}{4\sigma^2}\right).
\end{align}

The state that would remain invariant if an event occurs after the measurement of the position is the statistical mixture,
\begin{align}
\rho_2(0) &= \vert a\vert^2 \ket{\varphi_1}\!\bra{\varphi_1} + \vert b\vert^2 \ket{\varphi_2}\!\bra{\varphi_2}.
\end{align}

We can distinguish both states computing the expectation value of the observable $O_S=p$. \mbox{In~fact, while}
$$\mathrm{Tr}[p \rho_2(0)]=0,$$
we have that
\begin{align}
\tr{p \rho_1(0)} = - i\hbar (ab^* - a^*b)  \frac{L}{4\sigma^2}  \exp\left(-\frac{L^2}{8\sigma^2} \right).
\end{align}

Hence, if~the initial state is chosen with the appropriate phases, p discriminates whether an event occurred or not. 
While taking into account that fundamental uncertainties on the measurement of length intervals forbid a perfect preparation of the wavepackets $\vert \varphi_1 \rangle$ and $\vert \varphi_1\rangle$, since they imply errors $\Delta L$ and $\Delta \sigma$ on the separation and width of the Gaussians, which will lead to uncertainties on the expectation value of~$p$.

We have shown, in~\cite{GaGaPu18}, that a simple propagation of uncertainties on the error that is induced on 
${\rm Tr} [p \rho_1(0)] - {\rm Tr} [p \rho_2(0)]$  by these uncertainties gives,

\begin{equation}
    \Delta^2 \left( \tr{p \rho_{S,1}(0)} - \tr{p \rho_{S,2}(0)} \right)\ge 2  \tr{p \rho_{S,1}(0)}^2 \frac{\Delta L^2}{L^2},
\end{equation}

{The uncertainty} 
on the measurement of $O = p \otimes 1_E$, has to be taken into account when analyzing the global protocol that is considered above in Equation~(\ref{eq:observabledecay}).
\begin{equation}
\label{condition}
\tr{p \big( \rho_{S,E,1}(T_f)  -\rho_{S,E,2}(T_f)  \big) } 
 \le \Delta \left( \tr{p \left(\rho_{S,1}(0) - \rho_{S,2}(0) \right)} \right), 
\end{equation}

{That is}, 
 once  condition (\ref{condition}) is satisfied, the~uncertainty in the measurement of the observable prevents one from verifying whether the system was initially in a coherent superposition $\rho_{S,1}$ or in a statistical mixture $\rho_{S,2}$. 
Notice that this is a fundamental limitation and cannot be circumvented by making multiple measurements. It is related to the impossibility of preparing exactly the same initial state with infinite precision. One cannot decide whether the system was at the end of the evolution in a mixture state or the one that results from the evolution with the master equation. While, in~the case of ambient decoherence, the~final states were not distinguishable for any practical purpose (FAPP) when the fundamental decoherence due to the bounds in the precision of the determination of time and space intervals is taken into account; the production of events in a system that initially was in a pure state superposition is compatible with the evolution equation. This condition is fundamental and not FAPP. This determines when a system produces an event, and~it does so in an objective way. 
\mbox{In reference~\cite{GaGaPu18}}, we proved that events can occur for times $T>\tau_{\rm event}$, where,
\begin{align}
\label{eq:timeevent}
\tau_\ev =  \frac{ 1 }{ 2 (2\pi)^{3/2}  } \frac{ \tau_D^3 L^2}{ T_P^2 L_P^2}.
\end{align}

It could be argued that more efficient protocols could exist than the one that requires the system's time-reversal and, therefore, that the analysis presented is insufficient for showing that events can be produced without violating the causal evolution that is described by the master equation. Nevertheless, the~studies conducted by Brown and Susskind~\cite{BrSu} and Aaronson, Atia, and~Susskind~\cite{AaAtSu} regarding quantum complexity suggest that any protocol that allows for distinguishing the final states of a measurement process when the environment is included would be equally costly. In~particular, \mbox{it would} be as difficult as implementing a procedure for bringing the dead cat back to life, the~way in which the time reversal operation would do it and, therefore, would require operations with a similar degree of difficulty. On~the other hand, the~quantum complexity ideas that were mentioned before suggest that a rigorous demonstration of the indistinguishability of the final states after measurements, and~statistical mixtures can be~feasible.

\emph{{When the fundamental decoherence effects due to the use of real quantum and relativistic clocks and to the quantum and gravitational limits to the measurements of time intervals are taken into account, one can show that during the measurement process pure states evolve into states that are fundamentally physically indistinguishably from statistical mixtures of the various possible outcomes of the measured observable, without~any correction or violation of the evolution that is described by the master equation.}}

\section{The Production of Events and the Evolution of States during~Measurements}\label{sec5}

Generically, the~production of events alters the final state of a quantum system. The~only exception takes place when the events occur in statistical mixture states. In~that case, the~states can evolve according to the master equation without any change. In~the preceding sections, we have shown that when we take into account that time is not an ideal parameter, but~rather it is measured in clocks in a quantum gravitational universe, the~states evolve into statistical mixtures and, therefore, it is possible to explain why events take place in measurement~devices.

At this point, one could ask: ``isn’t the and/or problem still present?''. Have we adequately explained the transition from a superposition to a single outcome? Our point of view about this problem is that when the state of a system takes the form to be a statistical mixture \emph{{events occur}}. \mbox{This statement} could be considered as an ontological postulate, being the role of physics to predict when the events can occur and with which probability occur. When an event happens, like in the case of the dot on a photographic plate in the double slit experiment, typically many properties are actualized. For~instance, the~dot may be darker on one side than the other, or~it may have one of many possible shapes. The~postulated association between properties and objects that are typical of the classical physics is now substituted by an association of properties with events. Objects that are understood as systems in certain dispositional state do not have properties until they are measured or produce events. Observe that the production of events is not necessarily associated with a measurement. An~event occurs each time the state of a system becomes indistinguishable with a statistical~mixture.

All of the recent criticisms based on Wigner's friend, like that of Frauchiger and Renner~\cite{FrRe,FrRe-1} to quantum mechanics, stem from the assumption that the unitary evolution predicted by ordinary quantum mechanics with a classical time is correct with infinite precision. The~quantum decoherence due to the use of real clocks eliminates all of the problems mentioned before. Events can happen without violating the causal evolution described by the master equation, with~probabilities that coincide exactly with those that were predicted by the textbook~interpretations.

{\em {The treatment here presented is purely quantum mechanical, events and states provide a complete physical description. Events happen when the state acquires the form of a statistical mixture. One of the outcomes compatible with the statistical mixture is actualized. The~occurrence of an event does not require any change in the causal evolution of the statistical mixture. The~reduction of the state that is usually associated with measurement processes is nothing else than the update of the information resulting from the observation of \mbox{the state}.}}

\section{Conclusions}

We have presented a brief review of the Montevideo Interpretation of Quantum Mechanics while taking into account recent advances in the understanding of the problem of time~\cite{hoehn,hoehn-1} in general relativity and of quantum complexity~\cite{AaAtSu,BrSu}. We show that previous results not only still hold, but~are set on a more solid footing. The~key observation is to take seriously the notion that time has to be a physical observable and not a classical parameter. This induces a fundamental mechanism for the loss of coherence. Coupled to the fundamental limitations that are implied by general relativity and quantum mechanics, it leads to an objective notion of quantum event without reference to a classical world or observers. We showed that the mechanism that leads to the loss of coherence can be framed in terms of recent proposals to deal with the problem of time in quantum~gravity.

\vspace{6pt}
\authorcontributions{Authors contributed equally to this manuscript.}

\funding{ This work was supported in part by Grant 
NSF-PHY-1903799, funds of the Hearne Institute for Theoretical 
Physics, CCT-LSU,  Pedeciba and Fondo Clemente Estable 
FCE\_1\_2019\_1\_155865.}

\acknowledgments{We wish to thank Philipp H\"ohn for many useful comments.}

\conflictsofinterest{The authors declare no conflict of interest.}

\reftitle{References}


\begin{thebibliography}{999}
\bibitem{decoherence}
Bacciagaluppi, G.  The Role of Decoherence in Quantum Mechanics. The Stanford Encyclopedia of Philosophy (Fall 2020 Edition). Zalta, E.N., Ed.  Available online: \url{https://plato.stanford.edu/archives/fall2020/entries/qm-decoherence/}  {(accessed on 23 November 2020).} 
\bibitem{quda} {Ollivier, H.; Poulin, D.; Zurek, W.}  
{Environment as a witness: Selective proliferation of information and emergence of objectivity in a quantum universe}.
  {Phys. Rev. A} \textbf{2005}, \emph{72}, 042113.

\bibitem{coarse} Kofler, J.; Brukner, C. {Classical World Arising out of Quantum Physics under the Restriction of Coarse-Grained Measurements}.  \emph{Phys. Rev. Lett.} \textbf{2007}, \emph{99}, 180403. 

\bibitem{givingup}
Saunders, S.; Barrett, J.; Kent, A.; Wallace, D. {Many Worlds? Everett, Quantum Theory and Reality}. Oxford University Press: Oxford, UK, 2010.

\bibitem{givingup-1}
Everett, H., III. {"Relative State" Formulation of Quantum Mechanics}. \emph{Rev. Mod. Phys.} \textbf{1957}, \emph{29}, 454. 

\bibitem{givingup-2}
DeWitt, B. {Quantum mechanics and reality}. \emph{Physics Today} \textbf{1970}, \emph{23}, 30.
\bibitem{givingup-3}
Vaidman, L. Many-Worlds Interpretation of Quantum Mechanics. The Stanford Encyclopedia of Philosophy (Fall 2018 Edition). Zalta, E.N., Ed.  Available online: \url{https://plato.stanford.edu/archives/fall2018/entries/qm-manyworlds/} {(accessed on 23 November 2020).} 

\bibitem{objectivity}
Fuchs, C.; Schack, R. {Quantum-Bayesian coherence}. \emph{Rev. Mod. Phys.} \textbf{2013}, \emph{85}, 1693. 

\bibitem{objectivity-1}
Fuchs, C.; Mermin, D.; Schack, R. {An Introduction to QBism
with an Application to the Locality of Quantum Mechanics}. {Am. J. Phys.} \textbf{2014}, \emph{82}, 749. 

\bibitem{singleworld}
Ghirardi, G.C.; Rimini, A.; Weber, T. {Unified dynamics for microscopic and macroscopic systems}. \emph{Phys. Rev. D} \textbf{1986}, \emph{34}, 470. 

\bibitem{singleworld-1}
Di\'osi, L. {Gravitation and quantum-mechanical localization of macro-objects}. \emph{Phys. Lett. A} \textbf{1984}, \emph{105}, 199. 

\bibitem{singleworld-2}
Di\'osi, L. {Continuous quantum measurement and Itô formalism}. \emph{Phys. Lett. A} \textbf{1988}, \emph{A129}, 419.

\bibitem{singleworld-3}
  Pearle, P. {Combining stochastic dynamical state-vector
reduction with spontaneous localization}. \emph{Phys. Rev. A} \textbf{1989},
39, 2277.

\bibitem{singleworld-4}
Ghirardi, G.C.; Pearle, P.; Rimini, A. {
Markov processes in Hilbert space and continuous spontaneous localization of systems of identical particles}. \emph{Phys. Rev. A} \textbf{1990}, \emph{42}, 78--89.

\bibitem{singleworld-5}
Gisin, N. {Stochastic quantum dynamics and relativity}. \emph{Helv. Phys. Acta} \textbf{1989}, \emph{62}, 363--371.

\bibitem{singleworld-6}
Penrose, R. {On Gravity's role in Quantum State Reduction}. \emph{Gen. Rel. Grav.} \textbf{1996}, \emph{28}, 581.

\bibitem{allof}
Mueller, M.P. {Law without law: From observer states to physics via algorithmic information theory}.  \emph{Quantum} \textbf{2020}, 4, 301.


\bibitem{montevideo}
Gambini, R.; Pullin, J.
The Montevideo Interpretation of Quantum Mechanics: A Short Review.
\emph{Entropy} \textbf{2018}, \emph{20}, 413.


\bibitem{montevideo-1}
Gambini, R.; Garcia-Pintos, L.P.; Pullin, J.
An axiomatic formulation of the Montevideo interpretation of quantum mechanics.
\emph{Stud. Hist. Phil. Sci. B} \textbf{2011}, {\emph{42}}, 256--263.


\bibitem{montevideo-2}
Gambini, R.; Pullin, J.
The Montevideo interpretation of quantum mechanics: Frequently asked questions.
\mbox{\emph{J. Phys. Conf. Ser.}} \textbf{2009}, {174}, 012003.


\bibitem{DiracBergmann} Bergmann, P.
\emph{Rev. Mod. Phys.} \textbf{1960}, \emph{33}, 510. 
Observables in General Relativity
\bibitem{DiracBergmann-1}
Dirac, P.A.M.  \emph{Lectures on Quantum Mechanics};  Dover: New York, NY, USA, 2001.

\bibitem{butterfield} 
Butterfield, J. 
Assessing the Montevideo Interpretation of Quantum Mechanics.
\emph{Stud. Hist. Phil. Sci. B} \textbf{2015}, {\emph{52}}, 75--85.
doi:10.1016/j.shpsb.2014.04.001

\bibitem{hoehn}H\"ohn, P.A.; 
Smith, A.R.H.; Lock, M.P.E. 
The Trinity of Relational Quantum Dynamics. \emph{arXiv} \textbf{2019},
arXiv:1912.00033.

\bibitem{hoehn-1}
Hoehn, P.A.; Smith, A.R.H.; Lock, M.P.E.
Equivalence of approaches to relational quantum dynamics in relativistic settings. \emph{arXiv} \textbf{2020},
arXiv:2007.00580.

\bibitem{rovelli}
Rovelli, C.  \emph{Conceptual Problems in Quantum Gravity}; Ashtekar, A., Stachel, J., Eds.; Birkhauser: Boston, MA, USA, 1991; p. 126. 

\bibitem{rovelli-1}
  Quantum mechanics without time: A model
\emph{Phys. Rev. D} \textbf{1990}, \emph{42}, 2638.

\bibitem{rovelli-2}
  Time in quantum gravity: An hypothesis
\emph{Phys. Rev. D} \textbf{1991}, \emph{43}, 442.

\bibitem{dittrich}
  Dittrich, B.
  Partial and complete observables for Hamiltonian constrained systems.
  \emph{Gen. Rel. Grav.} \textbf{2007}, \emph{39}, 1891. 

\bibitem{dittrich-1}
Partial and complete observables for canonical general relativity. 
  \emph{Class. Quan. Grav.} \textbf{2006}, \emph{23}, 6155.

\bibitem{dittrich-2}
  Dittrich, B.; Tambornino, J.
A perturbative approach to Dirac observables and their spacetime algebra.
  \emph{Class. Quan. Grav.} \textbf{2007}, \emph{24}, 757.

\bibitem{pawo}
Page, D.N.; Wootters, W.K.
Evolution without Evolution: Dynamics Described by Stationary Observables.
\emph{Phys. Rev. D} \textbf{1983}, {\emph{27}}, 2885.


\bibitem{unwa}
  Unruh, W.G.; Wald, R.M.
  Time and the interpretation of canonical quantum gravity
\emph{Phys. Rev. D} \textbf{1989}, \emph{40}, 2598.

\bibitem{pedagogical}
Gambini, R.; Porto, R.; Pullin, J.
Fundamental decoherence from quantum gravity: A Pedagogical review.
\emph{Gen. Rel. Grav.} \textbf{2007}, {\emph{39}}, 1143--1156.


\bibitem{lindblad} Lindblad, G.
On the generators of quantum dynamical semigroups.
  \emph{Commun. Math. Phys.} \textbf{1976}, \emph{48}, 119.

\bibitem{several}
  Salecker, H.; Wigner, E.P.
Quantum Limitations of the Measurement of Space-Time Distances.
  \emph{Phys. Rev.} \textbf{1958}, \emph{109}, 571.

\bibitem{several-1}
  Karolyhazy, F.
Gravitation and quantum mechanics of macroscopic.
  \emph{Nuo. Cim. A} \textbf{1966}, \emph{42}, 390.

\bibitem{several-2}
  Ng, Y.J.; van Dam, H.
Limit to space-time measurement.
  \emph{Mod. Phys. Lett. A} \textbf{1994}, \emph{9}, 335.

\bibitem{several-3}
  Limitation to Quantum Measurements of Space‐Time Distances.
\emph{Ann. N. Y. Acad. Sci.} \textbf{1995}, \emph{755}, 579.

\bibitem{several-4}
Amelino-Camelia, G.  Limits on the measurability of space-time distances in (the semiclassical approximation of) quantum gravity.
 \emph{Mod. Phys. Lett. A} \textbf{1994}, \emph{9}, 3415.

\bibitem{several-5}
  Ng, Y.J.; Lloyd, S.
Black hole computers.
  \emph{Sci. Am.} \textbf{2004}, \emph{291}, 53.

\bibitem{several-6}
  Frenkel, A.
Dependence of the Time-Reading Process of the Salecker–Wigner Quantum Clock on the Size of the Clock.
  \emph{ Found. Phys.} \textbf{2015}, \emph{45}, 1561.

\bibitem{several-7}
Singh, T.P.
Quantum gravity, minimum length, and holography. \emph{arXiv} \textbf{2019},
arXiv:1910.06350.

\bibitem{bounds} 
Gambini, R.; Pullin, J.
Fundamental bound for time measurements and minimum uncertainty clocks.
\emph{J. Phys. Comm.} \textbf{2020}, {\emph{4}}, 065008.


\bibitem{curved} 
Fuente, D.d.; Romero, A.; Torres, P.J.
Uniform circular motion in general relativity: Existence and extendibility of the trajectories.
\emph{Class. Quant. Grav.} \textbf{2017}, {\emph{34}}, 125016.


\bibitem{spatial}
Gambini, R.; Porto, R.A.; Pullin, J.
Fundamental spatiotemporal decoherence: A Key to solving the conceptual problems of black holes, cosmology and quantum mechanics.
\emph{Int. J. Mod. Phys. D} \textbf{2006}, {\emph{15}}, 2181--2186.


\bibitem{GaGaPu}
Gambini, R.; Garcia-Pintos, L.P.; Pullin, J.
Undecidability as solution to the problem of measurement: Fundamental criterion for the production of events.
\emph{Int. J. Mod. Phys. D} \textbf{2011}, {\emph{20}}, 909--918


\bibitem{despagnat}
d'Espagnat, B.  \emph{Veiled Reality}; Westviey Press: New York, NY, USA, 2003.
\bibitem{GaGaPu18}

  Gambini, R.; Garc\'\i{}a-Pintos, L.P.; Pullin, J.
Single-world consistent interpretation of quantum mechanics from fundamental time and length uncertainties.
\emph{Phys. Rev. A} \textbf{2019}, {\emph{100}}, 012113.


\bibitem{AaAtSu}
Aaronson, S.; Atia, Y.; Susskind, L.
On the Hardness of Detecting Macroscopic Superpositions. \emph{arXiv} \textbf{2020}, 
arXiv:2009.07450.

\bibitem{navascues} Navascués, M.
Resetting Uncontrolled Quantum Systems.
  \emph{Phys. Rev. X} \textbf{2018}, \emph{8}, 031008. 

\bibitem{BrSu}
Brown, A.R.; Susskind, L.; Zhao, Y.
Quantum Complexity and Negative Curvature.
\emph{Phys. Rev. D} \textbf{2017}, {\emph{95}}, 045010.


\bibitem{FrRe} Frauchiger, D.; Renner, R.
Quantum theory cannot consistently describe the use of itself
  \emph{Nat. Commun.} \textbf{2018}, \emph{9}, 3711.

\bibitem{FrRe-1}
Bub, J.  What is really there in the quantum world? In \emph{Philosophers Look at Quantum Mechanics};
Springer: \mbox{New York}, NY, USA, 2019.






 
 

\end{thebibliography}
\end{document}